# Who Benefits from AI? Self-Selection, Skill Gap, and the Hidden Costs of AI Feedback

Christoph Riedl*
D'Amore-McKim School of Business, Northeastern University, Boston MA, USA
c.riedl@northeastern.edu

Eric Bogert
D'Amore-McKim School of Business, Northeastern University, Boston MA, USA
e.bogert@northeastern.edu

* Corresponding author

Feedback from artificial intelligence (AI) is increasingly easy to access and research has already established that people learn from it. But individuals choose when and how to seek such feedback, and more engaged and motivated individuals may seek it more, creating an illusion of effectiveness that masks self-selection. We investigate how the endogenous choice to seek AI feedback shapes both individual learning and collective outcomes. Using data from over five years and 52,000 individuals on an online chess platform, we show that motivated and higher-skilled individuals self-select into AI feedback use—and use it more productively. This self-selection creates an illusion of AI effectiveness: apparent learning gains disappear once endogenous motivation is accounted for. This same selection mechanism drives two population-level consequences. Because motivated, higher-skilled individuals benefit disproportionately, AI access widens the skill gap. And because individuals exposed to centralized AI feedback converge on common input from a centralized AI source, intellectual diversity declines. Leveraging 42 platform-level natural experiments, we show this diversity reduction is causal. Self-selection into AI use thus connects individual-level learning dynamics to collective-level consequences—a micro-macro linkage with implications for organizational learning, human capital development, and the design of AI-augmented work.

*Keywords.* AI, learning, future of work, endogenous feedback seeking, skill gap, intellectual diversity.

## 1. Introduction

AI is changing how we work.[1] Scholars and practitioners are intrigued by the possibility that AI could not only increase performance in the moment of use but help improve performance long-term through organizational learning and human capital development (Helfat et al., 2023; Joseph and Sengul, 2024; Puranam, 2021; Tong et al., 2021). Indeed, research has already documented instances of successful learning from AI (Brynjolfsson, Li, and Raymond, 2025; Choi *et al.*, 2025a, 2025b; Gaessler and Piezunka, 2023; Shin *et al.*, 2023) and many have voiced the hope that learning from AI feedback could level the playing field by taking existing skill differentials out of the equation (SimanTov-Nachlieli, 2025). However, a long history of research on both feedback seeking (Anseel *et al.*, 2015; Ashford, Blatt, and VandeWalle, 2003) and technology use in organizations (Griffith, 1996) suggests that engagement with AI for feedback seeking is not simply a function of access or ease of use, but is driven by individual motivation. That is, the decision to seek AI feedback and how deeply to engage with it is endogenous to individual characteristics and context. Furthermore, we know people respond to AI input with unusual heterogeneity (Glikson and Woolley, 2020; Logg, Minson, and Moore, 2019). If engagement with AI is endogenous, then studying AI's effects under forced exposure—as much of the experimental literature on AI does (Brynjolfsson *et al.*, 2025; Dell'Acqua *et al.*, 2026; Lee and Chung, 2024)—identifies the causal effect of AI use on performance but leaves open the question of what happens when organizations actually deploy these systems, people choose for themselves, and population-level outcomes depend on who selects in. In this paper, we address the research question of how individuals engage with AI feedback—who seeks it and when—and how this individual-level variation shapes population-level consequences for two outcomes central to organizational learning: the skill distribution and intellectual diversity.

We study these questions in a setting that offers a rare combination: widespread voluntary access to high-quality AI feedback, granular observability of individual decisions, and a large population over an extended period. Specifically, we use data from an online chess platform

[1] Among the many speculations about how, and how severely, AI will affect the labor market, Jerome Powell, chair of the US Federal Reserve, expects AI will make 'significant changes' to the economy and labor market. https://www.banking.senate.gov/hearings/06/18/2025/the-semiannual-monetary-policy-report-to-the-congress

(lichess.org) spanning over five years and 52,000 individuals. The platform offers a feature where individuals can use AI to retrospectively analyze completed games. During this analysis, they receive feedback from a highly accurate AI about the quality of each move they made, how much advantage they gained (or lost) with each move, and what the best move in each situation would have been. This AI analysis is extremely popular: 15% of games are analyzed and 29% of individuals analyze at least one of their first 50 games. While this AI feature is available for free to everyone, it is up to each individual to decide whether—and for which game—to seek AI feedback. Indeed, usage of the AI feedback is heterogeneous: the modal user analyzes no games at all, the average is 1.2 with a standard deviation of 4.2, and the most AI active individual analyzed 49 of their first 50 games. Our setting focuses on this endogenous choice to seek AI feedback. Our analysis links individual-level adoption decisions to population-level consequences for intellectual diversity and skill distribution, connecting the micro and macro levels of the organizational learning process. We first explore who seeks AI feedback and in which situations (success vs. failure), focusing specifically on endogenous selection to seek feedback using a control function approach. We then explore the consequences of endogenous selection on the skill distribution and intellectual diversity. Here, we leverage 42 major platform updates as natural experiments to explore whether AI feedback causally changes the intellectual diversity on the population level.

We find that motivated individuals self-select into AI use, creating an appearance of AI effectiveness. While individuals who use AI feedback appear to improve, these gains are entirely driven by endogenous motivation and effort. Once we account for this selection, AI feedback has no average causal effect on learning. Yet AI does not leave the population unchanged. Higher-skilled individuals benefit disproportionately from access to AI because they use it more often and more productively. As a result of this behavioral pattern, AI increases—rather than decreases—the skill gap in the population. Furthermore, we find that AI feedback causally reduces intellectual diversity. Because individuals draw on the same centralized AI, they converge on common strategies and narrow their repertoires, reducing the intellectual diversity of the population. Instead of leading to durable changes in human capital from skill development, AI feedback appears to amplify existing inequalities with additional collective downsides by homogenizing intellectual perspectives.

We make two key contributions. First, we contribute to the growing conversation on how AI reshapes work and organizing (Bailey *et al.*, 2019; Krakowski, Luger, and Raisch, 2023; Raisch and Krakowski, 2021). While prior research has established positive performance improvements from AI it has focused on experiments with exogenous variation in access to AI and strong incentives (Chen and Chan, 2024; Dell'Acqua et al., 2023; Kawaguchi, 2021; Noy and Zhang, 2023), leaving open questions about endogenous decisions when and how to engage with AI. We show that self-selection into AI use is endogenous, masking much of the performance gain previously attributed to AI. More importantly, when selection into AI use is correlated with pre-existing advantages, even a technology with no causal effect on individual performance can still compound and shape organizational inequality. This complements prior work focused on displacement, automation, and differential access (Autor, 2015; Brynjolfsson and Mitchell, 2017). Second, we extend the organizational learning literature which has long recognized the tension between individual and organizational learning (March, 1991; Simon, 1991) by drawing a connection between AI-mediated individual optimization and degrading collective intellectual diversity. This identifies a new micro-level mechanism for the exploration-exploitation tradeoff (Lavie, Stettner, and Tushman, 2010; Lazer and Friedman, 2007) in organizational learning—one that operates through coupled individual learning trajectories through shared algorithmic signals, rather than organizational-level routines or strategy.

## 2. Theory and Research Questions

Research on feedback seeking and learning from feedback has a long tradition in organization research (Ashford *et al.*, 2003; Brown and Duguid, 1991; Coutifaris and Grant, 2022; Riedl and Seidel, 2018). Feedback provides a crucial information resource for individuals in organizations to achieve goals and achieve mastery (Ashford, 1986). Feedback seeking is the conscious devotion of effort to determine the correctness and adequacy of behavior for attaining valued goals (Ashford and Cummings, 1983). Research suggests that individuals carefully evaluate the costs and benefits of asking for feedback considering aspects such as the availability of feedback, the cost of feedback, and also emotional responses of what to do with the feedback once it is received (Anseel *et al.*, 2015; Ashford *et al.*, 2003).

Several differences stand out when considering seeking feedback from AI compared to from other humans such as friends, peers, or bosses. First, once deployed, AI systems can easily scale (Bengio *et al.*, 2024). Whereas a human manager or domain expert may be able to provide detailed feedback on only a few situations for a few members of the organization, AI systems have the capacity to provide detailed feedback to more individuals in more situations. That is, AI feedback is instantly available in many situations. Second, as AI systems approach expert-level quality, they can provide feedback to even the highest-skilled individuals who may otherwise face difficulty in getting access to elite mentors and coaches. Third, compared to asking humans for feedback, AI feedback comes without social cost such as concerns to appear confident and self-assured (Ashford, 1986). Individuals may feel higher psychological safety asking for corrective feedback from an AI due to the absence of negative social consequences (Edmondson, 1999). These differences make AI a fundamentally different source of feedback and patterns suggested by decades of research on human-to-human feedback cannot simply be assumed to hold.

These features point to an important difference between AI and other technologies studied in prior implementation research. Past work has correctly recognized that motivation to use a technology is crucial for adoption and that ease of use alone is insufficient (Griffith, 1996). However, AI feedback may alter the nature of this motivational challenge. Because AI feedback is instantly available, socially costless, and cognitively undemanding to access, the barrier to use is low—indeed, engaging with AI feedback may feel productive and affirming in itself. As a result, a new type of feedback seeking arises: one uninhibited by vanity that loses its developmental character; instead it encourages uncritical engagement in simple mental replay and retrospective indulgence in prior wins, instead of trying to understand reasons for failure. Indeed, recent experimental evidence suggests that when individuals have access to AI, they consult it on the majority of occasions in ways that bypass, replace, or reshape their intuition and deliberation (Shaw and Nave, 2026). Critically, this tendency toward passive reliance suggests that when AI feedback is freely available, the critical source of variation shifts from whether individuals seek feedback to how they engage with it—a distinction we term "access-effort decoupling" and develop in the following section. The deeper motivational challenge lies not in whether individuals use AI feedback, but in whether they engage with it productively: applying it to difficult situations, confronting corrective signals after failures, and investing the cognitive effort required to translate feedback into durable learning. This distinction is central to understanding why widespread access to AI feedback may not produce the learning gains that its availability would suggest.

Research has extensively studied who seeks feedback, why, and in which situations (see Ashford, Blatt, and VandeWalle, 2003 for a review). Skill and motivation are established predictors of feedback-seeking (Anseel *et al.*, 2015), and both shape engagement with AI feedback through the same basic structure: each predicts not only whether and how deeply an individual engages with feedback but also the trajectory of performance improvement independent of that feedback. More motivated individuals may seek AI feedback more often but would also improve more through their own effort; higher-skilled individuals engage with feedback differently but also perform better regardless of whether they receive it. Moreover, skill and motivation are not only parallel in the role they play but also empirically entangled through competence-motivation spirals (Salanova, Bakker, and Llorens, 2006) in which expertise and intrinsic motivation mutually reinforce each other. This is precisely why studying learning from AI under exogenous exposure can obscure its real-world consequences: the same individual differences that determine who engages with AI also determine who would have improved without it.

## 2.1 Why Learning from AI Feedback May Be Endogenous

We theorize three layers on which self-selection (i.e., endogeneity) may affect learning from AI feedback: (a) whether individuals seek AI feedback at all, (b) in which situations they seek AI feedback, and (c) how deeply they engage with the AI feedback.

***Seeking AI feedback at all.*** Feedback seeking, when the advisor is human, tends to be concentrated among more motivated individuals. One reason for this is goal importance (Wanberg and Kammeyer-Mueller, 2000). The higher the importance of the goal to the performer, the more frequently the performer seeks feedback (Ashford, 1986). Task identification also suggests that individuals who identify more with a task and value it more highly also tend to be more highly motivated (Deci and Ryan, 2000). This can lead to competence-motivation spirals where expertise development and intrinsic motivation mutually reinforce each other over time, further motivating those who already are highly motivated and skilled to exert effort and seek resources—such as AI feedback—to improve even more (Salanova *et al.*, 2006). Since AI feedback comes without associated social costs, individuals may also feel higher psychological safety asking for corrective feedback from an AI due to the absence of negative social consequences (Edmondson, 1999). This could alleviate image concerns and increase the desire to seek AI feedback, especially for higher-skilled individuals who tend to be more concerned about their image and to appear confident and self-assured (Ashford, 1986). Taken together, the availability of AI feedback alone may not be sufficient as feedback seeking is not automatic and may require high levels of motivation, and may correlate with skill.

***Type of AI feedback being sought.*** Motivation may not only affect feedback seeking overall, but also in which situations individuals seek AI feedback. This applies specifically to the distinction between positive feedback after successes compared to negative feedback after failures. Individuals generally try to defend and protect their egos (Baumeister, 2010). This generates a motive to avoid negative feedback (Ashford and Cummings, 1983; Wood, 1989). The tendency to avoid negative feedback may be particularly strong when the feedback is actively solicited (rather than provided without asking) as it may be more difficult to disregard once it is received (Ashford *et al.*, 2003). Even in situations in which honest feedback is crucial to achieve one's goals, people strongly favor receiving positive feedback that reinforces their positive self-perception, rather than negative feedback that challenges it. Consequently, individuals carefully choose situations that expose the self to positive feedback (i.e., successes) and use a variety of mental strategies to avoid feedback that threatens their positive self-image (i.e., failures; Ashford and Cummings, 1983; Baumeister, 2010). High levels of motivation—which result from strong identification with the task (Deci and Ryan, 2000) or epistemic curiosity (Metcalfe, Schwartz, and Eich, 2020)—may be necessary to seek feedback after failures. Beyond motivation, skill independently shapes in which situations individuals seek feedback. Higher skilled individuals who identify more with the task are also likely to have higher task-related self-esteem (Stets and Burke, 2000) and may thus be more likely to seek feedback in challenging situations. High self-esteem performers are resilient, have great confidence reserves, are more promotion focused (Lanaj et al., 2012), and therefore, can “take” negative feedback and will be more likely to seek feedback, even after experiencing failures.

***Deep engagement with AI feedback.*** Seeking feedback alone is not enough to guarantee positive results. The link between feedback seeking and positive performance gains is surprisingly weak in general (Anseel *et al.*, 2015). Engaging with the feedback more deeply, for example through reflection, increases its effectiveness (Anseel, Lievens, and Schollaert, 2009). Deep reflection, however, depends on both motivation and cognitive scaffolding. Individuals low in learning goal orientation and for whom the task is less important tend to engage less in reflection (Anseel *et al.*, 2009). And while people have tried to leverage AI to become more reflective (Abdel-Karim et al., 2023), promoting human reflection through AI seems difficult to trigger (e.g., Ma et al., 2024). Beyond motivation, individuals also require the necessary supportive cognitive structures for AI feedback to be effective. For example, Jussupow *et al.* (2021) show that only individuals who developed AI interrogation practices were able to benefit from AI. This can be especially important when AI judgements diverge from initial human judgements without AI ability to provide explanations. This can increase the recipient's uncertainty (Lebovitz, Lifshitz-Assaf, and Levina, 2022) and thus potentially lower future performance. In our empirical analyses, we can instrument for the first two (whether someone uses AI, and whether they use it after losses), but deep engagement is something we theorize about rather than directly measure.

Taken together, these three sources of endogenous selection suggest “access-effort decoupling”: observed correlations between AI feedback seeking and performance improvement may reflect pre-existing individual differences rather than causal effects of AI.

## 2.2 Collective Consequences of AI Feedback

The individual-level dynamics described above do not remain confined to the individual level. A central insight of organizational theory is that individual-level behavior often aggregate into collective patterns that no individual intended or anticipated (Levinthal and March, 1993; Sproull and Kiesler, 1991). When many individuals independently interact with the same centralized AI system, two collective consequences may emerge: a widening of the skill distribution, and a narrowing of intellectual diversity. Both arise not from AI's direct effects on any single individual, but from the accumulation of endogenous individual-level choices across a population.

***AI Effect on Skill Gap.*** A long-standing question is whether AI will increase or decrease the existing skill gap between higher- and lower-skilled workers (and as a result related outcomes like income inequality; Furman and Seamans, 2019). Two competing theories persist. On the one hand, lower-skilled decision-makers may benefit most. AI could act as an equalizer that helps close the skill gap. This is because the marginal impact of the quality of advice is decreasing in the ability of the individual who receives it (Chade and Eeckhout, 2018). Several empirical studies report that lower-skilled workers benefited more from access to AI than higher skilled (Choi and Schwarcz, 2024; Dell'Acqua *et al.*, 2026; Kanazawa *et al.*, 2022; Noy and Zhang, 2023; Riedl and Weidmann, 2026). In several of these studies, as a result, inequality between workers decreased. While the marginal impact logic has been applied mostly to AI's direct effect on performance, it may also hold for learning. Intuitively, since learning increases at a decreasing rate (Allen and Choudhury, 2022; Becker, 1962; Foster and Rosenzweig, 1995; Mithas and Krishnan, 2008), learning from AI should improve the skill of lower-skilled individuals faster and could thus help close the skill gap and reduce inequality.

On the other hand, research has also shown that this may not happen in practice. Skill and AI often act as complements (Acemoglu and Autor, 2011; Autor, 2024; Autor, Levy, and Murnane, 2003). Drawing benefits from AI may depend on the user's ability to correctly judge quality of AI input (Agrawal, Gans, and Goldfarb, 2022) and request AI input in the right situations (see section above on feedback seeking). Learning from AI may itself depend on relevant prior experience (Brynjolfsson *et al.*, 2025). Understanding when and how to best use AI systems may constitute an important skill itself (Riedl and Weidmann, 2026). For example, Bouacida et al. (2024) find that lower-skilled individuals often fail to heed the advice they receive due to overconfidence or intrinsic preference. Prior theory has suggested that inexperienced individuals suffer even more from their limited understanding of AI outputs which would make it harder to leverage AI assistance in the right way and learn from it (Kellogg, Valentine, and Christin, 2020). Similarly, Wang et al. (2023) show that workers with more task-based experience benefit more from AI. That is, the ability to learn from AI may itself be a valuable skill of high practical importance as small differences in competence at learning tend to accumulate (Levitt and March, 1988). Understanding whether access to AI feedback increases or decreases the skill gap is crucial to predict downstream labor-market effects (Eloundou *et al.*, 2023; Kim *et al.*, 2024), and address concerns about AI induced inequality and human de-skilling (Hendrycks, Mazeika, and Woodside, 2023; Raisch and Krakowski, 2021; Russell, 2022).

***Diversity loss through convergent updating and specialization.*** When many individuals consult the same centralized source of information, their beliefs and strategies tend to converge toward a common set of outputs, replacing previously heterogeneous private knowledge with shared knowledge derived from that source (Sourati, Ziabari, and Dehghani, 2026; Wenger and Kenett, 2025). This homogenization risk is especially acute when individuals have access to the same feedback from a single centralized AI system (such as AI foundation models, or super-human chess AI). Indeed, AI induced cognitive homogenization has already been reported (Doshi and Hauser, 2024; Kleinberg and Raghavan, 2021; Meincke, Nave, and Terwiesch, 2025). The population-level convergence may be further amplified if individuals who rely on AI feedback also narrow their exploration. This is plausible because specialization—focusing on a narrow area of expertise—is a well-recognized consequence of learning (Cohen, Levinthal, and others, 1990; Levinthal and March, 1993). Together, convergent updating and specialization produce homogeneity in mental patterns. This population-level loss of intellectual diversity could then negatively affect long-term problem solving ability of firms (Levinthal and March, 1993; Page, 2019) and undermine firms' competitive advantage (Felin and Holweg, 2024). This poses a crucial alignment problem and is a recurring theme in work discussing unintended consequences and societal level risks of AI (Bengio *et al.*, 2024; Bommasani *et al.*, 2022).

## 2.3 Research Questions

The arguments above raise several interrelated research questions, which we examine empirically. We investigate the effect of AI feedback on two levels of analysis: individual vs. collective.

First: On the individual level, who seeks AI feedback and why? Beyond documenting observed behavioral differences, we are in particular concerned with endogenous variation in motivation that could explain engagement with AI (overall, but also in specific situations such as successes vs. failures). Hence, we explore this question "in the field" in which AI is merely available but individuals are free to decide when and how to use it, and where other modes of learning are also available. Empirically, access-effort decoupling should manifest in measurable outcomes: high prevalence of AI feedback seeking overall, particularly after successes, but with limited positive effects on performance once accounting for endogenous selection.

Second: On the collective level, we ask whether endogenous selection into AI use produces unintended downstream consequences. Specifically, does access to AI feedback increase or decrease the skill gap over time, and how does it affect long-term intellectual diversity?

## 3. Setting and Data

We use the adoption of AI in chess as a sample research domain. Chess is a particularly relevant context to study AI because of its long history and tradition of using AI. This setting offers three features that make it particularly well suited to studying endogenous selection into AI use. First, AI is broadly available and its high-quality is well known (Kasparov, 2020), meaning that variation in adoption reflects individual choice rather than differences in perceived value or trust in AI (Glikson and Woolley, 2020). Second, adoption is fully voluntary, isolating selection dynamics from compliance or mandate effects common in organizational IT implementations. Third, performance has no practical ceiling and even elite participants benefit substantially from AI feedback. This rules out the possibility that non-adoption reflects rational disengagement by individuals who have little left to gain.

Our data come from lichess.org, a popular and free online chess platform. On average, more than 90 million games are played each month on the platform.[2] It is distinct among the major chess platforms because it is a nonprofit, and makes every platform feature available for free.[3] Our study centers around a specific platform feature: The platform allows players to seek AI feedback on games they played on the platform. This AI feedback is powered by a chess AI called Stockfish, the strongest chess AI in the world.[4] That is, the AI feedback players receive can be considered feedback from a chess expert of the highest caliber. The current (in 2025) version is Stockfish 16, and has a chess skill (called Elo) of 3,641 (see more on the chess Elo skill measure further below). For comparison, if Magnus Carlsen, widely regarded as the best player in history, played against this AI at his peak rating (Elo = 2,882), he would be expected to win just 12 out of 1,000 games.

[2] See https://database.lichess.org/#standard_games for the full database of Lichess games

[3] Competitor platforms such as Chess.com offer similar AI features but only behind paywalls.

[4] https://chessify.me/blog/top-chess-engines

The AI feedback informs players of the strength of every move they played, and the strategic advantage they gained (or lost) with each move. The AI feedback also informs players what the best moves would have been in each position and how much these alternative moves would have strengthened their position. The AI feedback feature allows decision-makers to "replay" a game, move by move, and receive AI feedback on every move (Figure A1 in the Appendix shows an annotated example of the AI feedback). We refer to the process of having an AI analyze a game as AI feedback and AI analysis interchangeably.

We then exploit a unique feature of the Lichess platform: Lichess tracks and displays whether a given game has been analyzed using the AI feedback feature. We collected this data to investigate individual-level performance returns to seeking AI feedback on games. Unfortunately, the platform does not show *who* analyzed the game. To overcome this limitation we focus on games played by humans against bots. For those games we can be reasonably certain that if a human played against a bot, the human player in the match was the person who analyzed the game. Technically, other players could also seek such AI feedback but since games against AI bots are predominantly played for training purposes (Gaessler and Piezunka, 2023), the sheer number of those games, and that only a few of these games are played by famous players, this is highly unlikely.

Consequently, the starting point of our data collection are AI bots. Lichess hosts AI-powered bots of varying skill levels that players can challenge for practice. Playing against bots is primarily a training activity; players select bots matched to their skill level to practice openings, tactics, and endgame technique. Bots are clearly labeled as such on the platform — they do not masquerade as human players. We identified every AI bot on Lichess in September 2022 (n = 129). We then collect every game played by each bot using Lichess' API. We had 1,054,640 games after removing games between AI bots (with no human player), games that were not "normal" chess games (e.g., 4 player games) and unrated games. We removed unrated games because players likely used more effort in their rated games. Finally, we included every game up to the human player's 50th game. We chose this cutoff because only 5% percent of players played more than 50 games. This cutoff increases the generalizability of our results, as they are less influenced by a few players who have played unusually many games. On average, it took 53 days for a player to play 50 games against a bot, with a standard deviation of 144 days. As a robustness check, we look at cutoffs of up to a player's 30th game and up to a player's 70th game and find substantially the same results (see Table A1 in the Appendix).

The final dataset contains 403,010 game-level observations from 52,251 human players. For each game, we have information on whether the game was analyzed (i.e., whether the human player sought AI feedback for the game), the performance of the human player in the game, and several other game- and individual-level attributes (see more details below). Notice that the number of observations in the different regression models will deviate from this number for two reasons. First, by using lagged variables on the right hand side, the first observation for every individual will drop out (e.g., the first game observed by an individual has no meaningful count of prior AI feedback). Second, when we use logistic regression, instances with no variation in the dependent variable are collinear with the individual-level fixed effect and cannot be used for estimation (e.g., individuals who either analyzed all games or no games).

***Dependent Variables.*** Our study relies on two key dependent variables: AI feedback, and performance. AI feedback is a simple dummy indicator which is 1 if a human player sought AI feedback on a given game, and 0 otherwise. To capture human performance we construct a measure of overall move accuracy. First, for each game we collect the entire move-by-move sequence of play from the Lichess' API. We then use Stockfish 16 to analyze the quality of each move in every game in our sample (this corresponds to 70,037,446 moves total, or about 35 moves per game). To assess the quality of each individual move, Stockfish simulates up to 1,500,000 moves to assess what the best strategy is for each side (this is the same setting as the AI feedback feature on the Lichess website). Move quality is expressed as the probability that white or black would win the game after that move. Finally, we transform this move-level performance measure into an aggregate game-level performance measure using the same formula employed by Lichess.[5]

[5] https://lichess.org/page/accuracy

***Control Variables.*** We collected both game- and individual-level data as control variables. Game-level data includes the length[6] and type of the game (e.g., bullet or classic), whether the game was won or lost, whether the game used a common chess opening, and the skill of the bot opponent. Individual-level data includes the human player's skill level (measured through Elo, see Appendix Equation A2 for details), and tenure and experience of the human player (Table 1). We also measure both a player and a bot's skill using their Elo scores. An Elo score measures how good a player is, with higher numbers indicating more skill. Most bots have a consistent underlying skill that only changes when the bot is updated, which rarely happens, but the bot Elo still varies slightly based on games they have won and lost. Since moving first gives players more control over the direction of the game, we include an additional dummy indicating whether the focal player moved first (playing white) or second (playing black, omitted category) as an important control variable (the Section *Defining Control Variables* in the Appendix provides detailed explanation of each measure).

We include a control variable $Tenure_{it}$ and its squared term, which captures the number of years an individual *i* has been active on the platform at time *t*. This allows us to distinguish between learning from AI Feedback and changes in performance that occur because of other factors (Argote and Epple, 1990; Thornton and Thompson, 2001). $Tenure_{it}$ effectively captures the general passage of time and thus incorporates possible experience accumulated outside the platform through playing in-person games, games played on other platforms, other forms of chess training, education, or other activities outside the online platform (Riedl and Seidel, 2018). We also include a count of total games played against bots, which we call *experience*, because prior experience is a critical determinant of learning (Riedl and Seidel, 2018).

[6] 107 games in our dataset last longer than 100,000 seconds. These games are "correspondence" games, which are one of the six game types in our data. Correspondence games give each player a long time (usually one day) to make a move. This explains the very large maximum value on game length in Table 1.

Table 1. Descriptive statistics and correlations of key study variables.

| | Mean | SD | Min | Max | (1) | (2) | (3) | (4) | (5) | (6) | (7) | (8) |
|---|---|---|---|---|---|---|---|---|---|---|---|---|
| AI Feedback (1) | 0.15 | 0.36 | 0.00 | 1.00 | | | | | | | | |
| Performance (Move Accuracy) (2) | 87.98 | 6.49 | 7.79 | 100.00 | 0.12*** | | | | | | | |
| Opponent Skill (3) | 1693.17 | 323.17 | 543.00 | 3039.00 | 0.00 | 0.07*** | | | | | | |
| Human Skill (4) | 1588.91 | 302.00 | 455.00 | 3409.00 | 0.04*** | 0.22*** | 0.43*** | | | | | |
| Experience (5) | 14.27 | 13.40 | 1.00 | 49.00 | 0.02*** | 0.01** | -0.07*** | -0.04*** | | | | |
| Tenure (6) | 0.15 | 0.36 | 0.00 | 4.68 | 0.04*** | 0.01*** | -0.01*** | 0.06*** | 0.23*** | | | |
| Common Opening (7) | 0.84 | 0.37 | 0.00 | 1.00 | 0.02*** | 0.03*** | -0.01*** | -0.04*** | 0.01*** | 0.01*** | | |
| Game Length (8) | 810.98 | 37716.42 | 0.00 | 7683335.00 | 0.01*** | 0.01*** | 0.01*** | 0.01*** | 0.00 | 0.02*** | 0.00* | |
| Loss (9) | 0.57 | 0.49 | 0.00 | 1.00 | -0.16*** | -0.38*** | 0.30*** | -0.11*** | -0.06*** | -0.02*** | 0.00* | 0.00 |

# 4. Analysis

## 4.1 Seeking AI Feedback

### 4.1.1 Empirical Model

We investigated *who* seeks AI feedback and in *which* situations with logistic panel regression. The dependent variable is whether an individual requested AI feedback on a given game (1 = analyzed, 0 = otherwise). We estimate:

$$Pr(AI\ Feedback_{it} = 1) = logit^{-1}(\beta_1 Loss_{it} + \beta_2 Skill_{it} + \beta_3 Loss_{it} \times Skill_{it} + controls$$

$$\alpha_i + \alpha_{bot} + \alpha_{year} + \alpha_{month} + \alpha_{GameType} + \epsilon_{it})$$

where $Pr(AI\ Feedback_{it} = 1)$ is the probability that individual *i* seeks AI feedback for the game played at time *t*. $Loss_{it}$ is an indicator whether the game to be analyzed was a win or a loss, $Skill_{it}$ is the player *i*'s Elo rating at time *t*, and *controls* (experience, tenure, tenure squared, common opening, game length). Finally, α are fixed effects for the individual, bot, year, month, and game type.

***Endogeneity in Feedback-Seeking.*** To capture the endogenous selection to seek AI feedback, we use the control function approach (Wooldridge, 2010). Control function analysis is a form of instrumental variable analysis but it has the benefit that it can be easily applied to nonlinear models (binomial; Wooldridge, 2010). We construct individual-level instrumental variables following similar ideas as prior work in other settings of management research (van Angeren *et al.*, 2022; Kummer and Schulte, 2019). Specifically, we use four instruments to capture endogeneity due to time-variant omitted variables like motivation, engagement and effort: (1) the average number of losses in the five previous games of the individual (more losses could indicate higher ambition by choosing more challenging opponents to play against); (2) the time elapsed between the previous five games and the current game (shorter time span—playing more often—could indicate higher motivation); (3) a dummy indicator set to 1 if the player is an inexperienced user with fewer than five games on the platform (this could indicate lower awareness of the AI feedback feature); and (4) the strength of the opponent (also capturing ambition). The idea behind those instruments is that they capture time-varying levels of motivation, engagement, and effort which may influence losing the focal game and seeking AI feedback. However, this fluctuation in motivation only affects the likelihood to seek AI feedback on the focal game through the loss of the focal game (exogeneity). Our instruments are designed to capture time-varying factors such as motivation and engagement that jointly drive AI usage and performance. While the control function residual is by construction a composite of all unobserved confounders, our theoretical framework identifies intrinsic motivation as the most plausible driver among these. We provide more details on the estimation and the results of the first-stage regression in the Appendix.

#### 4.1.1 Results

We find individuals have a strong aversion to seeking AI feedback in situations of failure (i.e., for games they have lost; Model 1: $\beta = -1.61;\ p < 0.001$). There is only an 11% chance that an individual will seek AI feedback on a failure while there is a 21% likelihood to seek AI feedback in case of success (average marginal predictions). Furthermore, higher-skilled individuals are significantly more likely to seek AI feedback than lower-skilled individuals (Table 2, Model 1, $\beta = 0.06;\ p = 0.031$). Going up one standard deviation in skill decreases the odds of seeking AI feedback by about 2%, and for losses, the odds increase by about 23%. Furthermore, we find substantial heterogeneity across skill: higher-skilled individuals are much more likely to seek AI feedback in situations of failure (large and significant coefficient of the interaction term $\beta = 0.24;\ p < 0.001$ in Model 2). Looking at the endogeneity corrected results in Models (3) and (4) we indeed find signs of endogeneity: a substantially large and significant coefficient for the control function residual from the first stage. This suggests that the strong negative relationship between seeking AI in situations of failure is driven in large part by endogenous time-varying shocks (e.g., lower motivation). We find signs of a negative bias, where time-variant shocks that promote losses (e.g., lower motivation) are correlated with lower likelihood to seek AI feedback. Accounting for this endogeneity we now find a significant *positive* coefficient for analyzing losses (Model 3: $\beta = 0.14$; $p = 0.004$). That is, after accounting for self-selection, individuals appear to be aware of the benefit and potential for learning from failures (Riedl and Seidel, 2018). The coefficient for the interaction between skill and loss is virtually identical (Model 4: $\beta = 0.23$; $p < 0.001$).

In summary, we find substantial evidence of endogeneity in seeking AI feedback on losses, as the coefficient changes markedly after controlling for the first-stage residual. However, the interaction between loss and skill is robust to this correction, indicating that the endogeneity mechanism does not systematically vary with skill. This suggests that while self-selection biases the estimated mean effect of seeking feedback on losses, they do not affect how this impact differs by skill level. The key insight is that higher-skilled and more motivated, engaged individuals are more likely to seek AI feedback on losses.

Table 2. Propensity to seek AI feedback. Models (1) and (2) show plain binomial regression, Models (3) and (4) show endogeneity corrected control function results.

| Dependent Variable: | Seek AI Feedback | | | |
|---|---|---|---|---|
| | Binomial | | Endogeneity Corrected | |
| | (1) | (2) | (3) | (4) |
| Loss | −1.62*** | −1.59*** | 0.14** | −0.06*** |
| | (0.03) | (0.03) | (1.27) | (0.33) |
| Loss × Skill | | 0.24*** | | 0.23*** |
| | | (0.02) | | (0.02) |
| Controls | | | | |
| Skill (Elo, z-scored) | 0.06* | −0.04 | 0.08*** | −0.02*** |
| | (0.03) | (0.03) | (0.02) | (0.02) |
| Experience (log) | −0.00 | −0.00 | −0.00*** | 0.00*** |
| | (0.02) | (0.02) | (0.01) | (0.01) |
| Tenure | 0.39 | 0.37 | 0.42 | 0.40 |
| | (0.58) | (0.58) | (0.72) | (0.46) |
| Tenure$^2$ | 0.02 | 0.02 | −0.03*** | −0.02*** |
| | (0.06) | (0.06) | (0.09) | (0.05) |
| Common Opening | −0.01 | −0.01 | 0.02*** | 0.02*** |
| | (0.03) | (0.03) | (0.03) | (0.02) |
| Game Length | 0.00 | 0.00 | 0.00*** | 0.00*** |
| | (0.00) | (0.00) | (0.00) | (0.00) |
| Control Function: Loss | | | −1.78*** | −1.55*** |
| | | | (1.27) | (0.33) |
| Num. obs. | 176,884 | 176,884 | 176,884 | 176,884 |
| Num. groups: Individual | 8,304 | 8,304 | 8,304 | 8,304 |
| Num. groups: Bot | 57 | 57 | 57 | 57 |
| Num. groups: Year | 6 | 6 | 6 | 6 |
| Num. groups: Month | 12 | 12 | 12 | 12 |
| Num. groups: Game Type | 6 | 6 | 6 | 6 |
| Deviance | 130,085.00 | 129,818.25 | 129,974.54 | 129,735.50 |
| Log Likelihood | −65,042.50 | −64,909.12 | −64,987.27 | −64,867.75 |

$^{***}p < 0.001$; $^{**}p < 0.01$; $^{*}p < 0.05$; $^{\dagger}p < 0.1$

*Notes.* Models (1) and (2) show standard errors clustered on the individual level in parentheses, Models (3) and (4) show bootstrapped standard errors (1,000 replications) and *p*-values based on quantile bootstrap method.

## 4.2 Learning from AI Feedback

### 4.2.1 Empirical Model

In order to determine whether people improve as a result of using AI feedback, we use panel regression with user and time fixed effects. Like our prior model, this allows us to control for unchanging but unobserved individual qualities, such as demographics, ability, or number of games played prior to arrival on Lichess.org. We estimate the equation

$$Performance_{it} = \beta_1 AI\ Feedback_{it-1} + \beta_2 log(CumulativeGames_{it-1}) +$$

$$\beta_3 Tenure_{it} + \beta_4 Tenure_{it}^{2} + \beta_5 Opponent\ Skill_{it} + controls +$$

$$\alpha_i + \alpha_{bot} + \alpha_{year} + \alpha_{month} + \alpha_{GameType} + \epsilon_{it}$$

where *Performance*, *Tenure*, and *Opponent Skill* are indexed by the game played at time *t* and individual *i*. *AI Feedback* and *Cumulative Games* are time lagged—indexed by time *t-1*—to capture learning from *previous* AI feedback and direct experience (cumulative games). We also estimate variations of this equation in which we split the total amount of AI feedback into separate counters for AI feedback on wins and losses. As before, α are fixed effects for the individual, bot, year, month, and game type.

***Endogeneity in Feedback-Seeking.*** As in the previous analyses, we are especially concerned about the endogeneity in seeking AI feedback overall in addition to seeking it in specific situations (losses). Motivated individuals are more likely to become aware of the AI feedback feature, and also more likely to use it. That motivation and effort could then also be the driver behind their performance, not the AI feedback per se. That is, individuals learn because they are motivated, not because they use AI feedback. To test if AI feedback itself is necessary—or whether motivated individuals would have learned the same amount without AI feedback—we need a shock that affects availability of the AI feature even for motivated individuals. Whereas the AI analysis is in principle always available, for free, to every user on Lichess and nominal access is universal, the effective access may vary in ways that act like constraints. We focus on two specific constraints: knowledge of the feature and varying availability/usability of the AI feature due to server outages and maintenance windows. Specifically, we use the same control function approach as before and estimate an additional first-stage model and then include the residual from it in the main regression. For this model, we use AI feedback seeking as the dependent variable and pick two new instruments. First, awareness of the AI feature itself may govern its use. We define a dummy variable set to 0 if the individual has never used the AI analysis before, and 1 after their first use. Before learning about the AI analysis feature, usage is impossible or very unlikely, giving us exogenous non-access. After discovering the feature, individuals could use, if they want and endogenous usage depends on motivation. Since awareness is mechanically correlated with learning progress it is important to explicitly account for confounding time trends; we hence include controls for tenure, tenure squared, and total games played to soak up practice intensity and career trajectory effects. Second, we include a supply-side instrument for the availability of the AI feature by counting the total number of games analyzed with AI by other players on the same day. This instrument complements the previous one, as it relies on platform-wide variation rather than individual choice. To further strengthen exogeneity we include day-of-week and month controls to capture common time trends & seasonality (e.g., total number of AI analyses by others may spike when more people are online such as weekends). Together, the two instruments strengthen identification and robustness as they employ different logics: the awareness instrument relies on within-player knowledge shocks whereas the "usage by others" instrument relies on global usage thus capturing supply-side availability shocks. These two instruments give us plausible exogenous variation in whether a player can or will use the AI feature. This will help us to more precisely answer the question whether motivation causes people to use AI feedback and said AI feedback is necessary for them to learn. This helps us more precisely identify the causal benefit of AI feedback by shedding light on the counterfactual that the same motivated individuals would have achieved the same learning by other means (e.g., by using other platform features like opening drills and puzzles).

#### 4.2.1 Results

We find no effect of past AI feedback on current performance (Table 3; Model 1; $\beta = 0.00;\ p = 0.552$). However, this overall null-effect hides two opposing heterogeneous effects. Whereas feedback on losses leads to performance improvements (Model 2; $\beta = 0.05;\ p < 0.001$), feedback on wins actually deteriorates performance ($\beta = -0.05;\ p < 0.001$). AI feedback on losses helps, AI feedback on wins hurts performance. These results hold also when considering performance against human players (Appendix Table A3). That is, AI feedback is useful for broad-based learning that improves performance both against AI and human opponents. Model 3 & 4 shows the endogeneity corrected result (Table A2 in the Appendix shows the first-stage regressions). We find large and significant effects for both control function residuals ($\beta = -5.17$; $p < 0.001$ and $\beta = 0.33$; $p < 0.001$). The significant residuals imply that there is endogenous self-selection and that our instruments explain enough variation in AI use to detect it. The residual for access is positive and significant, showing that unobserved factors that increase access also raise learning, as expected: more motivated or engaged individuals seek out AI feedback more often and see higher performance increase. The residual for AI use in productive situations (losses) is negative and significant, suggesting that unexplained variation in productive use—beyond what is predicted by access and other instruments—is negatively correlated with performance. This likely reflects that players who overuse AI feedback in inefficient ways, or use it reactively when struggling, do not benefit from AI; in fact, AI may even hinder their performance. The structural features of the estimation help recover a proxy for motivation and effort. Once we control for these endogenous channels, the direct causal effect of all forms of AI feedback are zero (total AI feedback, AI feedback on losses, AI feedback on wins).

Table 3. Effect of prior AI feedback on current performance. Standard errors clustered on the individual level in parentheses.

| Dependent Variable: | Performance | | | |
|---|---|---|---|---|
| | OLS | | Endogeneity Corrected | |
| | (1) | (2) | (3) | (4) |
| Total AI Feedback | −0.00 | | 0.00*** | |
| | (0.00) | | (0.00) | |
| AI Feedback on Losses | | 0.05*** | | −0.00*** |
| | | (0.01) | | (0.01) |
| AI Feedback on Wins | | −0.05*** | | −0.00*** |
| | | (0.01) | | (0.01) |
| Controls | | | | |
| Skill (Elo, z-scored) | 0.17*** | 0.19*** | 0.17*** | 0.17*** |
| | (0.04) | (0.04) | (0.03) | (0.03) |
| Experience (log) | −0.01 | −0.00 | −0.02*** | −0.01*** |
| | (0.03) | (0.03) | (0.02) | (0.02) |
| Tenure | −0.82 | −0.77 | −0.51 | −0.51 |
| | (0.92) | (0.92) | (0.68) | (0.68) |
| $\text{Tenure}^2$ | −0.27** | −0.27** | −0.26*** | −0.26*** |
| | (0.09) | (0.09) | (0.07) | (0.07) |
| Common Opening | 0.10* | 0.10* | 0.10*** | 0.10*** |
| | (0.05) | (0.05) | (0.04) | (0.04) |
| Game Length | 0.00* | 0.00* | 0.00*** | 0.00*** |
| | (0.00) | (0.00) | (0.00) | (0.00) |
| Control Function: Loss | | | −5.17*** | −5.17*** |
| | | | (0.03) | (0.03) |
| Control Function: AI Analysis | | | 0.33*** | 0.33*** |
| | | | (0.05) | (0.05) |
| Num. obs. | 176,543 | 176,543 | 176,543 | 176,543 |
| Adj. $R^2$ | 0.18 | 0.18 | 0.18 | 0.18 |
| Num. groups: Individual | 8,304 | 8,304 | 8,304 | 8,304 |
| Num. groups: Bot | 58 | 58 | 58 | 58 |
| Num. groups: Year | 6 | 6 | 6 | 6 |
| Num. groups: Month | 12 | 12 | 12 | 12 |
| Num. groups: Game Type | 6 | 6 | 6 | 6 |
| Num. groups: Day of Week | 7 | 7 | 7 | 7 |

$^{***}p < 0.001$; $^{**}p < 0.01$; $^{*}p < 0.05$; $^{\dagger}p < 0.1$

The significant residuals suggest that endogenous factors like motivation and effort strongly drive both the use of AI feedback and performance. Motivated individuals are more likely to use AI, and also more likely to use it in challenging situations (to analyze losses). These individuals would have learned just as much without AI feedback because the underlying driver for their performance is motivation and effort, not AI feedback per se. Our theoretical framework identifies a third layer of endogeneity—depth of engagement with feedback—that we theorize but cannot directly measure; to the extent that reflective engagement further separates productive from passive AI use, our estimates represent a conservative test of the role that endogenous motivation plays in observed AI learning gains. At the margin, given observed usage patterns, AI feedback has no independent causal effect on performance beyond individuals' own motivation. In other words, AI does not add value automatically. Merely offering access to AI does not help disengaged or less motivated individuals increase their performance and only works as a complement to effort and preexisting skill. Performance gains from using AI are endogenous to motivation or discipline; and conditional on motivation and effort, AI feedback does not improve performance.

### 4.3 The Population Skill Gap: How Individual Selection Reshapes Collective Distributions

The preceding analyses showed that AI feedback has no average causal effect on performance once endogenous selection is accounted for. However, the selection dynamics themselves are not uniform across the skill distribution—both in terms of how much individuals use AI feedback but also the situations in which they seek AI feedback (losses). What are the descriptive consequences of this non-uniform adoption? To characterize how AI access reshapes the skill distribution in practice we estimate heterogeneous treatment effects using a Generalized Random Forest (GRF). GRF is a non-parameteric, machine learning based method with automatic variable selection (Athey, Tibshirani, and Wager, 2019).

We find statistically significant and practically significant heterogeneity in the effect of AI feedback on performance (Figure 1). The effect of AI feedback is significantly higher for high-skilled individuals than for low-skilled individuals (based on median split: low = 0.021 [95% CI: 0.019 - 0.023] vs. high = 0.037 [95% CI: 0.034 - 0.040]). These effects do not account for endogeneity and thus reflect the combined effect of AI feedback and the correlated motivation and engagement patterns documented above. As such, they capture the realized relationship between AI use and performance improvement as it actually unfolds in the population, and it is this realized relationship—whether causal or not—that determines how the skill distribution evolves over time. This pattern could be driven by skilled individuals adopting AI feedback more (and more effectively) while less motivated or less skilled people do not fall behind. Whether AI feedback is uniquely enabling or merely one vehicle among many that motivated individuals would substitute away from if unavailable remains an open question. But in an environment where AI feedback is widely available, low-cost, and high-quality, it is the vehicle through which skill advantages currently compound.

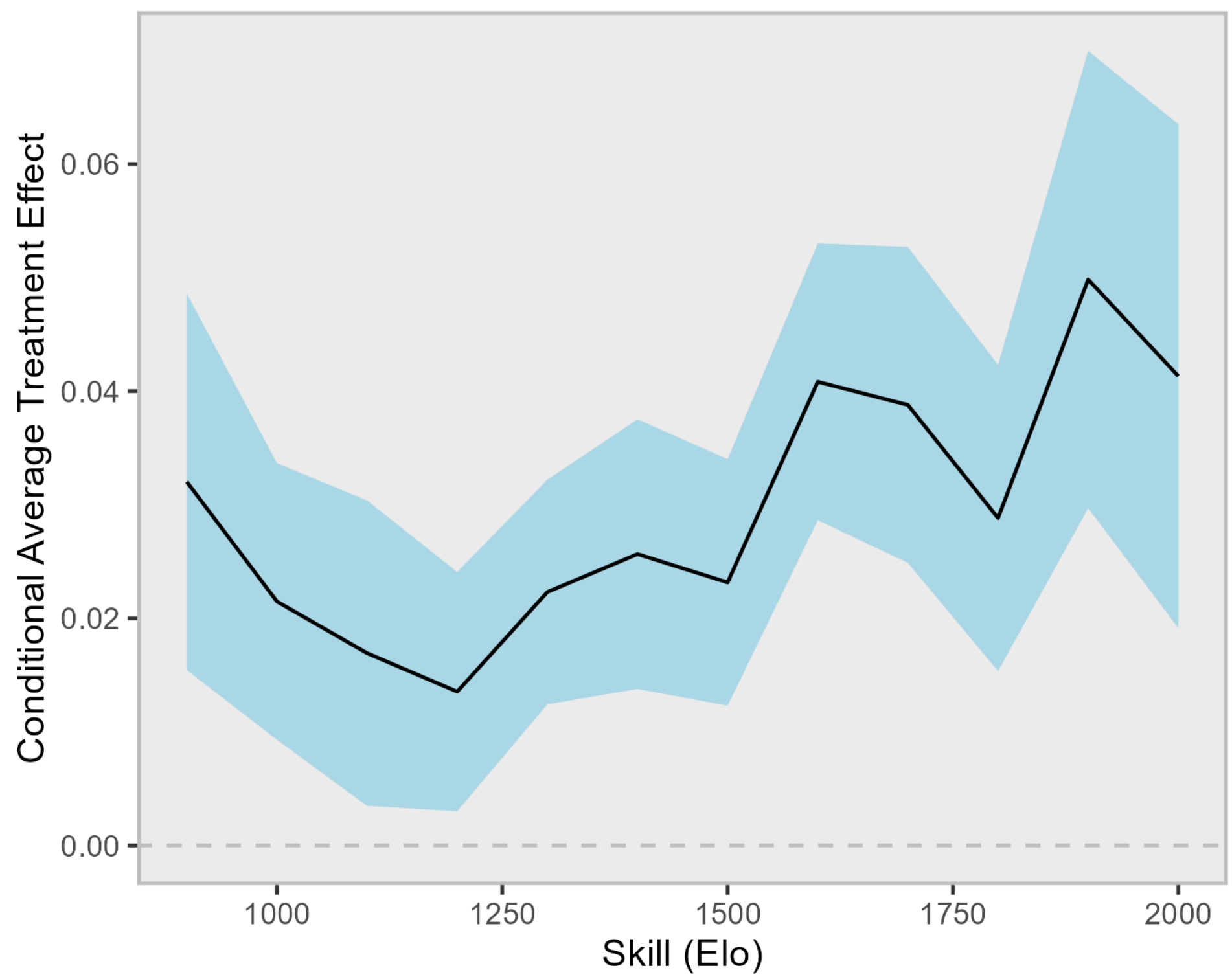

Figure 1. Across the population, broad availability of AI feedback tends to increase the skill gap as higher-skilled learn faster per instance of AI feedback than lower-skilled.

### 4.4 Population-Level Intellectual Diversity

To test our predictions about homogenizing effects of AI feedback on intellectual diversity at the population level, we focus on chess opening strategies. Openings are the initial move sequences of a game, and they are consequential because they define the strategic landscape for everything that follows. Many opening sequences have well-known names such as the “Queen's Gambit” and the “Sicilian Defense” (Hooper and Whyld, 1992). Specifically, we classify opening strategies using three character Encyclopaedia of Chess Openings (ECO) codes. Most players develop a repertoire of favored openings, and this repertoire reflects their strategic identity—a narrow repertoire signals deep specialization while a broad one signals flexibility (Chassy and Gobet, 2011). We begin with a simple descriptive population-level analysis that points to strong homogenization (Figure 2). We group individuals by their level of past AI feedback and compute diversity of opening moves as $1 - Gini$ coefficient. We find that many individuals appear to specialize along the same dimensions, thus reducing group-level intellectual diversity: As individuals accumulate more exposure to AI feedback, the diversity of opening move strategies they employ decreases ($\rho_p = -0.59$; $p = 4.9 \times 10^{-6}$).

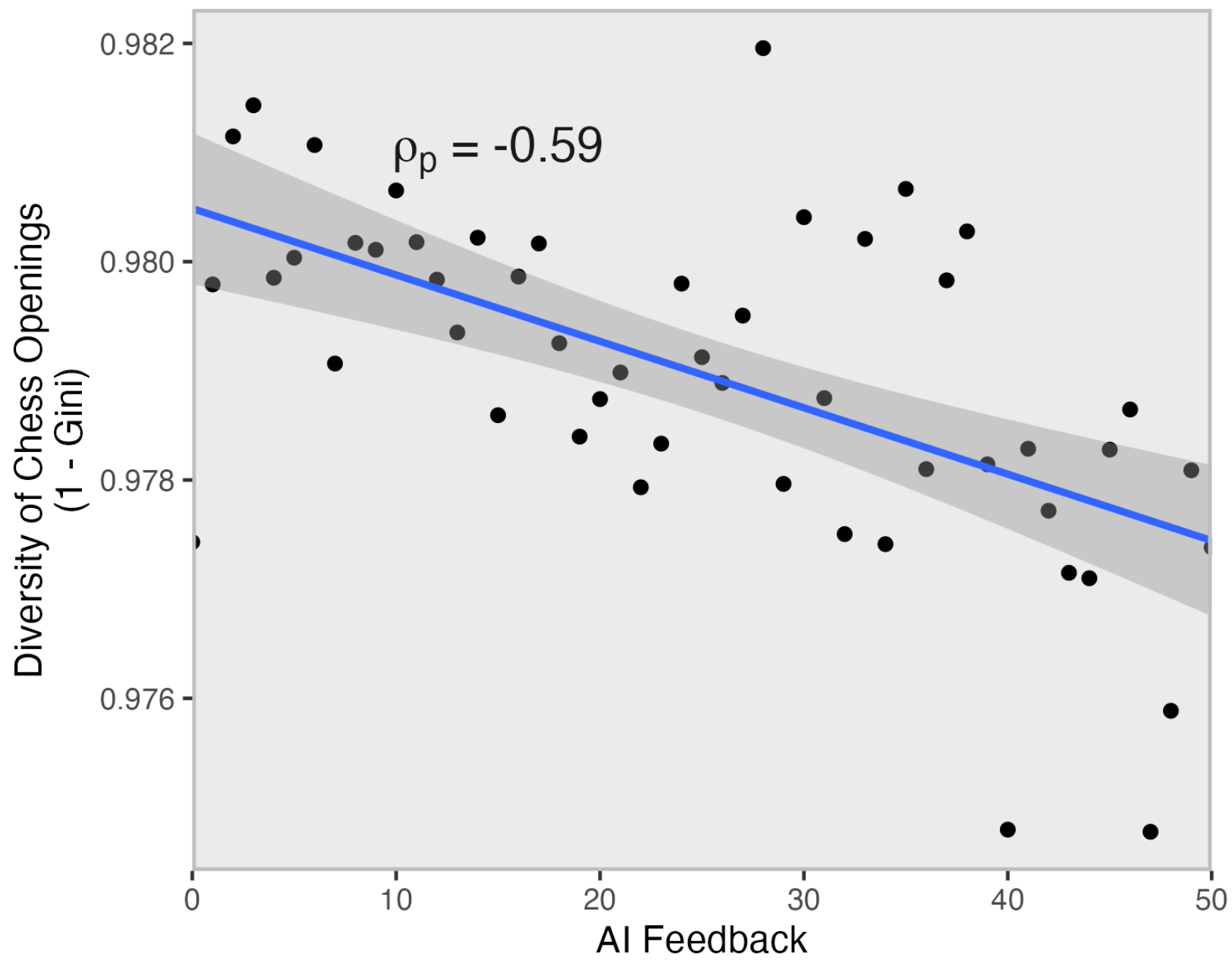

Figure 2. Collective intellectual diversity decreases with more AI analysis experience.

Next, we provide causal evidence for lower population-level intellectual diversity by analyzing a set of natural experiments. To explore the causal effect of AI feedback on population-level intellectual diversity we combine a Regression Discontinuity in Time (RDiT; Hausman and Rapson, 2018) analysis and natural experiments. First, we aggregate all chess openings played on the entire platform on a given day into a diversity metric ($1 - Gini$). This metric captures how likely it is that two games played on the same day and drawn at random use the same opening sequence.[7] This gives us a single time series of platform-level intellectual diversity. Next, we collected data on changes to the platform which we will leverage as natural experiments to establish a causal effect of the AI feedback feature. Between 2020 and 2023[8] the platform experienced 42 major updates.[9] We treat these updates as regression discontinuity events by comparing the platform-level time series of intellectual diversity across multiple small time windows around these 42 platform changes.

[7] To account for the fact that diversity of strategies may depend mechanically on the number of daily games we aggregate, we draw 100 random games, and compute the platform-level diversity index for this random subset. To minimize noise from this sampling procedure, we then repeat this process 100 times and average the diversity score.

Each time window compares intellectual diversity right before vs. right after the platform changes in a stacked difference-in-difference analysis (Baker, Larcker, and Wang, 2022). Stacked difference-in-differences addresses bias from heterogeneous treatment effects in staggered adoption settings by comparing each treated unit only to appropriate not-yet-treated or never-treated units, aligned in event time. This avoids the problematic weighting and contamination from already-treated units that can distort estimates in traditional two-way fixed effects models. Onto this RDiT analysis we layer a natural experiment lens: 24 platform changes affected the AI feedback feature while 18 were unrelated.[10] The unrelated features serve as an active benchmark capturing the effect of platform updates on intellectual diversity via pathways other than the AI feedback feature. That means in addition to the dummy variable indicating whether a time series observation was obtained before vs. after a platform change, we introduce a second dummy variable indicating whether the platform change was related to the AI feedback or not. This analysis then compares changes in daily platform-level strategy diversity from right before a platform change to that after the platform change while simultaneously accounting for the fact that some changes were related to the AI feedback feature and others were not. That is, days right before a platform change serve as control cases for days right after a platform change; and days after an AI unrelated platform change serve as controls for days after an AI related platform change. Finally, as platform-level strategy diversity may react to attention and salience from platform changes in general (Clarke *et al.*, 2023), we also sample 100 placebo events that were independent of any platform change events. This analysis identifies the local average treatment effect of changes to the AI feedback feature on strategy diversity.

[8] We focus on days with at least 100 games (the period from 2020-11-27 through 2023-06-09) and exclude early days of the platform on which only few daily games were played.

[9] https://lichess.org/changelog accessed 2024-07-08.

[10] The fact that we observe many platform changes of which some are related to the AI analysis feature and others are unrelated creates a form of "cross-sectional" variation which casts this analysis as a kind of difference-in-difference analysis (Angrist and Pischke, 2009). That is, we have many observations before and after platform changes, where in some the relationship between AI analysis should be unchanged from the before-to-after period (because the platform change was unrelated to AI analysis), in others it should be affected. We classify a platform change as AI related if the release notes for the main platform changes mention the word "analysis", and unrelated otherwise.

Specifically, we estimate

$$Diversity_{et} = \beta_1 After_{et} \times Placebo_e + \beta_2 After \times Unrelated + \beta_3 After_{et} \times AI\ Related_e + \alpha_e + \alpha_{et}$$

where $Diversity_{et}$ is the platform-level strategy diversity (z-scored) on day *t* around the natural experiment *e* and $\alpha_e$ are fixed effects for each natural experiment. This analysis estimates four levels of effects: Diversity in the period before the platform change (dropping out due to the fixed-effect formulation), after a placebo platform change, after a platform change that was unrelated to the AI analysis feature, and after a platform change related to the AI feedback feature. In a robustness test, we also include $Time$ and $Time^2$ controls (treating the entire estimate more as a continuous time regression discontinuity model rather than a plain regression discontinuity design without time as the running variable).

For each of the 42 platform change events (and placebo events), we select small and precise time windows of the 15 days prior to the platform change, and the 15 days after. This strengthens our Regression Discontinuity design with a mass of observations very close to the threshold where identification rests upon a conditional expectation as one approaches the threshold. Furthermore, since platform changes are unexpected for most users, individuals cannot easily sort across the threshold.[11] That is, we believe we can make a credible claim to exogeneity: Most individuals will be unaware when platform changes are being released, and maybe in particular be unaware whether the platform changes will affect the AI analysis feature or not. As a result, selection across the threshold and strategic behavior seem unlikely. We think this natural experiment setting approaches the as-good-as-random interpretation of a cross-sectional Regression

[11] We explicitly exclude platform changes related to so-called "community features" where individuals may have prior knowledge of the upcoming platform changes and could change their behavior in anticipation.

Discontinuity design. To account for correlation in residuals across events due to overlapping windows we use robust standard errors clustered by date.

***Results.*** We find AI related platform updates cause a significant decrease in intellectual diversity (Table 4: e.g., $\beta =- 0.14$; $p = 0.043$ in Model 1). Our placebo tests show that the RDiT design does not spuriously detect discontinuities at random dates (insignificant and near-zero coefficient across all models). On the other hand, platform changes unrelated to AI feedback systematically increase intellectual diversity. This is consistent with the effect of unrelated platform updates on general skill building and developing broad competence by enhancing the various tutor & puzzle features on the platform. These features are designed for practicing tactical motifs (pins, forks, skewers, mating nets, etc.) and pattern recognition by exposing players to diverse positions taken from real games. Thus, our main estimate should be interpreted as the additional effect of AI related platform updates relative to this typical upward jump associated with other platform updates. In net terms, AI related platform updates produce a decrease in intellectual diversity, overcoming the baseline tendency for platform updates to increase intellectual diversity. Results are robust to using 30 day windows before/after the platform change and including time controls.

Table 4. Causal effect of AI feedback on platform-level intellectual diversity. Robust standard errors in parentheses are clustered on the day level.

| Dependent Variable: | Intellectual Diversity (1 - Gini; OLS) | | | |
|---|---|---|---|---|
| | ±15 days window | | ±30 days window | |
| | (1) | (2) | (3) | (4) |
| After × Placebo | −0.06 | −0.03 | −0.04 | 0.00 |
| | (0.03) | (0.05) | (0.03) | (0.03) |
| After × AI Unrelated | 0.19** | 0.19** | 0.17* | 0.17* |
| | (0.07) | (0.07) | (0.07) | (0.07) |
| After × AI Related | −0.14* | −0.14* | −0.27*** | −0.27*** |
| | (0.07) | (0.07) | (0.06) | (0.06) |
| Controls | | | | |
| Time | | −0.00 | | −0.00 |
| | | (0.01) | | (0.00) |
| $Time^2$ | | −0.00 | | 0.00 |
| | | (0.00) | | (0.00) |
| Num. obs. | 4,260 | 4,260 | 8,520 | 8,520 |
| Num. Nat. Experiments | 142 | 142 | 142 | 142 |
| Adj. $R^2$ | 0.41 | 0.41 | 0.35 | 0.35 |

$^{***}p < 0.001$; $^{**}p < 0.01$; $^{*}p < 0.05$

***Narrowing Specialization on Individual-Level.*** What could explain this persistent population-level homogenization? We supplement the population-level analyses with additional mechanistic analysis on the individual level (see Appendix). Here, we find individuals narrow their repertoires with AI exposure—and skilled players do so even more—which amplifies the convergence effect. Specifically, we use a simple logistic regression framework to study individuals' likelihood to repeat the same opening strategy that they used in one of the previous five games as a function of how often they sought AI feedback in the past (see Appendix for method details and full results). We find that additional AI feedback significantly increases the likelihood that an individual repeats the same opening strategy (β = 0.04; $p$ = 0.03). This is not just happening for beginners who are new to the game. Instead, the rate of specialization is even higher for higher skilled individuals (significant positive interaction between skill and past AI feedback; β = 0.02; $p$ = 0.026).

## Discussion

Our analysis shows that, descriptively, individuals learn from AI feedback after failures but not after successes. These performance gains are, however, entirely driven by endogenous selection such as motivation and effort. The apparent benefits of AI feedback are largely or entirely attributable to the same unobserved factors that drive AI adoption and engagement. Learning is concentrated among higher-skill, highly motivated individuals, due to their endogenous tendency to seek more AI feedback and use it more productively (after a failure). This individual-level pattern has important implications on the population level: Instead of lifting lower-skilled individuals who have the most to gain, AI disproportionately complements higher-skilled individuals—amplifying the existing skill gap. This pattern is the empirical signature of access-effort decoupling: universal access to AI feedback lowers the barrier to use but not the barrier to learning, so the distribution of effort—not access—determines who gains. Furthermore, access to AI feedback causes a decrease in intellectual diversity due to convergent updating and individual-level specialization.

While prior research has established positive performance improvements from AI it has focused on experiments with exogenous variation in access to AI and strong incentives (Chen and Chan, 2024; Dell'Acqua *et al.*, 2026; Kawaguchi, 2021; Noy and Zhang, 2023), leaving open questions about broad-based and longer-term effects of AI where decisions to engage with AI are endogenous. Insights from our work complement two closely related papers that have indicated that people learn from AI but that did not directly observe who was using AI and why (Gaessler and Piezunka, 2023; Shin *et al.*, 2023). This pattern is not new. Apparent returns to new technologies are often driven by selection: agents that adopt are already more productive or better matched to the technology, so naive estimates overstate causal effects (e.g., Acemoglu et al., 2023; Suri, 2011). We extend this literature by pointing to endogenous selection—such as motivation and effort—as key hurdles for AI feedback to enhance competitive capabilities: engagement quality determines outcomes, not access alone.

Not only may selection mask apparent AI efficiency, it can also lead to a broader set of AI-driven inequalities. For example, substantial inequalities have emerged as women appear much less likely to adopt generative AI (Humlum and Vestergaard, 2024) and individuals with higher levels of education and income are more likely to adopt (United Nations Development Programme, 2025). This pattern is consistent with the competence-motivation spirals theorized above (Salanova *et al.*, 2006): higher-skilled individuals are more motivated to seek AI feedback, use it more productively after failures, and thereby widen the very skill advantage that fueled their engagement in the first place. This is consistent with cumulative advantage dynamics (Merton, 1968; DiPrete & Eirich, 2006) in which initial advantages in exploiting a new resource compound over time. Critically, our results suggest this process is driven not by ability alone but by the agency to seek and productively engage with AI feedback—echoing calls for research on how individual initiative generates, rather than merely correlates with, performance inequality (Aguinis & O'Boyle, 2014). As a result, broad-based positive equalizing effects from AI may be unlikely. This helps explain why AI may increase inequality over time, even if the short-term effects are neutral (Alderucci *et al.*, 2024) or benefit lower-skilled users more (Riedl and Weidmann, 2026). Those equalizing effects may be specific to settings focused on direct-AI-use-benefits, that are short-term, leverage direct performance incentives, and tasks with ceiling effects; they may be absent when learning and self-selection into (longer-term) productive use are more prominent. Noy & Zhang (2023), for example, used high-powered incentives with a performance-based bonus that would more than double the base pay ($10 base and up to $14 bonus pay) and Chang et al. (2025) paid a $0.10 bonus for *each* correct answer (on top of a $5 base pay). While intended to improve ecological validity, such direct, piece rate performance incentives are not common in most cognitive work settings which instead rely more on relational contracts (Baker, Gibbons, and Murphy, 2002).

Our findings on intellectual diversity complicate and qualify recent work on AI and human learning (Choi *et al.*, 2025a; Shin *et al.*, 2023). Where those studies document that AI exposure can improve *individual* performance (and even novelty), the opposite may be true when looking at the organization-level. Our diversity findings echo Choi et al. (2025b) and Meincke et al.'s (2025) observation that AI reduces diversity, but extend it by identifying a micro-level mechanism—convergent updating and specialization driven by feedback from a centralized AI—that produces this macro-level outcome. This helps connect insights on individual-level learning and specialization effects with organizational-level human capabilities (Krakowski *et al.*, 2023). While organizational learning theory has long recognized that learning itself can be self-limiting—driving organizations toward exploitation at the expense of exploration (March, 1991) and narrowing attention to proximate competencies (Levinthal and March, 1993)—these dynamics have traditionally operated through organizational-level routines, incentive structures, and local search processes. Our findings suggest that generic AI systems, by exposing diverse users to identical evaluative standards, introduce a novel micro-level pathway to this same macro-level outcome: one that operates not through organizational routines but through the coupled cognitive trajectories of individuals responding to shared algorithmic signals.

This also complements research on AI in organizational settings that has so far focused primarily on automation and productivity gains (Brynjolfsson and Mitchell, 2017; Choudhury et al., 2020; Raisch and Krakowski, 2021). Those settings typically treat AI as a technology that performs tasks. When AI instead augments human cognition it becomes entangled with the organization's knowledge base as humans and AI systems mutually adapt and influence each other over time (Riedl, Savage, and Zvelebilova, 2026). Our study serves as real-world evidence of population-level homogenization and unintended downstream consequences, an important AI risk that is so far poorly understood (Bengio *et al.*, 2024).

Both our skill gap and intellectual diversity findings highlight a paradox for organizational learning theory: positive individual-level effects can create unintended population-level consequences. This micro-macro dynamic—where individual choices about technology engagement aggregate into population-level consequences—suggests that organizational learning from AI cannot be understood by studying individual users in isolation. The decision to adopt and engage with AI is itself shaped by the organizational and motivational context, and the population-level consequences of that engagement depend on the distribution of motivation and skill across the workforce. Simply deploying AI tools does not constitute an organizational learning intervention; without attention to the selection processes governing engagement, organizations may observe illusory learning gains while the underlying skill distribution shifts in unintended ways. Our analysis points to motivation and effort as crucial ingredients for gains from AI to materialize. In many workplace settings, individuals are free to decide when and how to use AI. Not everyone may end up using AI feedback equally often or equally well. Using AI correctly itself may emerge as a valuable skill.

The collective consequences we document may generalize most directly. Generative AI is already widely used as a feedback source in knowledge work—users routinely ask large language models to evaluate and improve their writing, arguments, and analyses, receiving detailed, line-by-line suggestions not unlike the move-by-move feedback in our setting (Chen and Chan, 2024). And just as chess players receive feedback from a single AI system, the vast majority of such interactions are now routed through a small number of frontier models—ChatGPT, Claude, Gemini, ByteDance—encoding similar training data and optimization objectives. To the extent that millions of writers, analysts, and decision-makers are refining their thinking through the same few systems, our findings suggest that the concentration of cognitive influence in centralized AI deserves attention not only as a market structure question but as an epistemological one.

Our study is not without limitations. Our research provides evidence of how humans recruit AI and learn from AI feedback from just one domain (chess). Since there is only limited prior work on how humans interact with AI feedback and learn from it, it is difficult to say how similar or different our setting is from others (see above for some thoughts on how it may generalize to other settings). Furthermore, our results may not generalize to other settings given the (perceived) ability of AI right now. At least since IBM's DeepBlue AI beat Garry Kasparov (Kasparov, 2020), it is well known among chess players that AI performs at super-human level. So AI is both factually giving high quality feedback and people know that it is high quality. As such, aspects like trust in AI are likely to play a minor role in our setting while it may be more of an issue in other settings. While this may not be the case today in many AI application scenarios, our research gives insights as to what we can likely expect in potential *future* applications. Finally, an alternative interpretation of reduced diversity is that it simply reflects quality convergence — as individuals learn, they converge on objectively better strategies, and 'diversity' was partly noise from inferior play. While we cannot fully rule this out, our finding that diversity reduction is concentrated among AI-exposed individuals suggests it is not the full explanation.

Our research has several practical implications. Identifying that AI's apparent benefits are entirely explained by endogenous motivation can inform managerial decisions, especially since AI is very costly, widely promoted, and frequently assumed to be transformative. Our results cast doubt on the implicit assumption that AI improves the performance of workers just by being available. Motivated individuals enhance their performance, regardless of AI. Firms—and society more broadly—cannot assume that rolling out AI will lift everyone's skill level as simply providing AI does not substitute for motivation and skills in how to use them well. AI alone is unlikely to close the skill gap if not complemented by other interventions that address motivation and train individuals how to use AI productively. It also warns managers about the risk of unintended homogenization which can undermine competitiveness. For organizations, this implies that AI deployment without attention to who engages with it may not merely preserve existing inequality but actively reshape the performance distribution toward greater concentration. In summary, our findings suggest that in settings with more difficult tasks and without performance incentives that directly encourage AI use, AI is likely to act as an amplifier of existing skill and motivation, not as an equalizer.

# Online Appendix

## AI Feedback on the Lichess Platform

The focal variable in our study is how often individuals seek AI to analyze their games. The example below shows the output of what the AI reports when it analyzes a game between two of the best players on Lichess. We have annotated some of the most important aspects. Every gray bubble with text and a black outline was written by us. The arrow pointing from the knight to the bishop and the white arrow with the blue outline for the pawn on c2 are both part of the output Lichess gives to the player. These arrows change based on the move the player is currently analyzing.

Figure A1. Example output of Lichess' AI feedback.

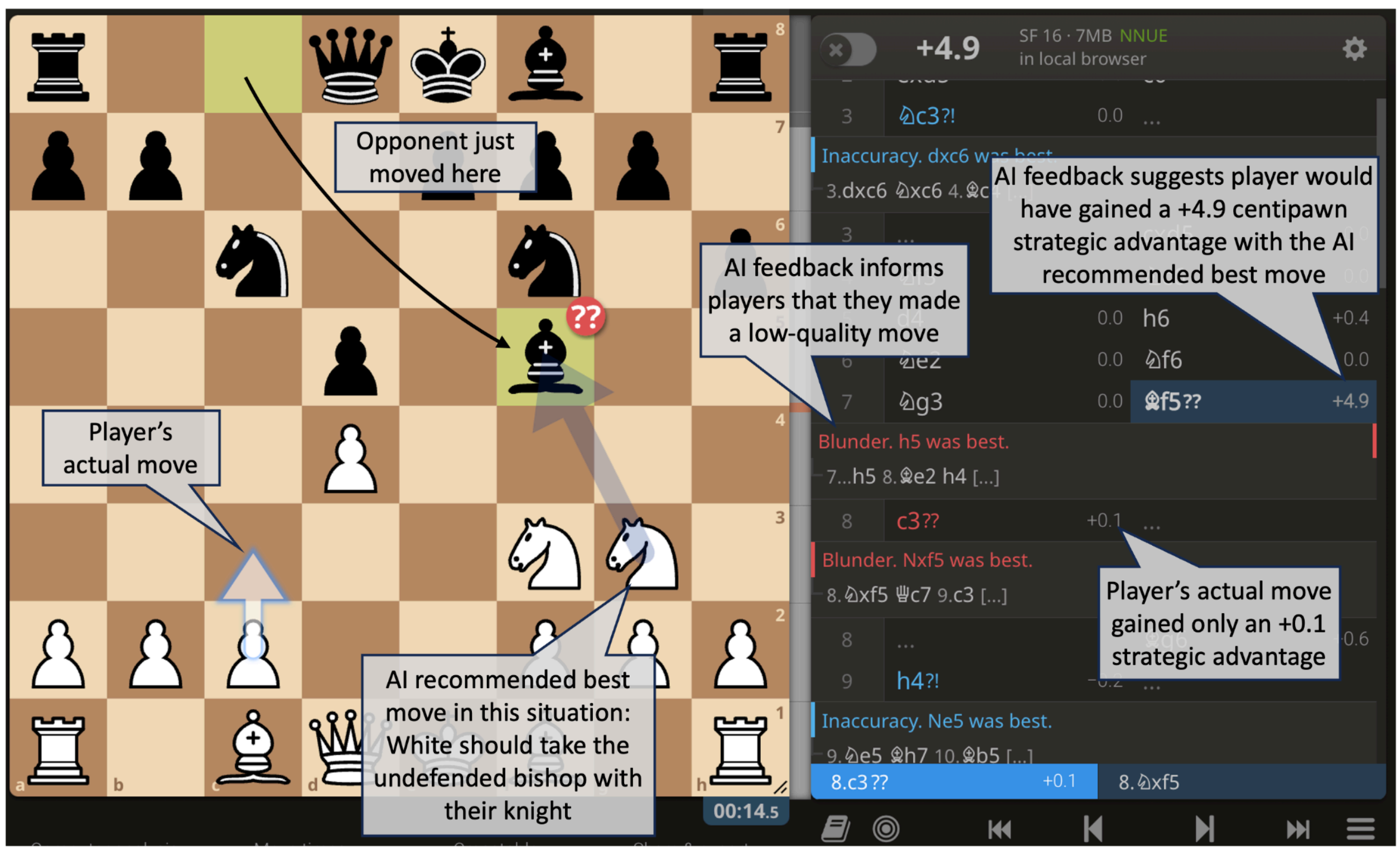

## Measuring Skill In Chess

One reason chess is studied so heavily is because a player's skill is accurately measured. Each player has a rating, usually called an Elo, that indicates how strong a player is (Elo 1978). Stronger players have higher ratings. Equation A2 below demonstrates the percentage of points that player A isplayer is expected to earn against player B. Thus, if player B's Elo is 200 points higher than player A's, then player B is expected to earn 24% of the points in a series of games (where a win is 1 point, a loss is 0 points, and a draw is 0.5 points).

Because Elo is a measure of skill, higher Elo players are expected to beat lower Elo players. In chess, all games conclude as either a win (1 point), a loss (0 points) or a draw (½ point). The percentage of points player A is expected to earn playing against player B is shown by the equation below.

Thus, if player A is 50 Elo higher than player B, player A is expected to win 57% of the available points in a match. If we exclude the possibility of draws, this means player A is expected to win 57% of the time.

$$E_A = \frac{1}{1 + 10^{\frac{Rating_B - Rating_A}{400}}} \qquad \text{Eq. A2}$$

## Defining Control Variables

We used characteristics of the human, the bot, and the game to identify the features of a game that are correlated with games being analyzed. The features included:

- The human’s skill (measured by Elo, then z-scored against all other player Elos that ever played against bots)
- A dummy variable for whether the human lost the game (1 = lost, 0 = won or drew)
- The interaction of the human’s skill and whether they lost

- Common Opening, a dummy variable indicating whether the game began with the two most common openings in chess - in which white's queen pawn or king pawn moves forward two spaces (1 = either of those moves, 0 = neither of those moves)
- Game Type describes the time control, where each player gets this amount of time for all moves. We treated these as dummy variables and incorporated them in the fixed effects
  - Ultrabullet: less than 30 seconds
  - Bullet: more than 30 seconds and less than 3 minutes
  - Blitz: More than 3 minutes and less than 8 minutes
  - Rapid: More than 8 minutes and less than 25 minutes
  - Classical: longer than 25 minutes
- Game Length, a continuous variable, measuring the number of seconds a game lasted
- Experience, a continuous variable, measuring the cumulative number of games a player has played against bots (logged)
- Tenure, a continuous variable, measuring the number of years the player had been on the platform at the time of the game
- $Tenure^2$, a continuous variable, measuring the number of years (squared) the player had been on the platform
- White: A binary variable indicating whether a player used the white pieces

## Robustness Test: Using Different Cutoffs of Games Played

Our models in the main paper used up to a player's 50th game. We used that cutoff because only 5% of players played more than 50 games. Including longer histories would imply that any effects we observe are driven by fewer and fewer individuals. For example, moving from histories of 50 games to 100 games, increases the number of observations in our dataset by 24%, but those additional observations come from only 5% of individuals making it harder to generalize to the overall population. The table below shows what the coefficients would be for Model 4, in Table 3 if we had used cutoffs at 30 games, 50 games, and 70 games.

Table A1. Estimation results from using cutoffs at 30 games, 50 games, and 70 games

| Dependent Variable: | Performance (OLS) | | |
|---|---|---|---|
| | First 30 Observations | First 50 Observations | First 70 Observations |
| AI Feedback on Losses | 0.11*** | 0.06*** | 0.03*** |
| | (0.01) | (0.01) | (0.01) |
| AI Feedback on Wins | −0.09*** | −0.04*** | −0.01*** |
| | (0.01) | (0.01) | (0.00) |
| Controls | | | |
| Experience (log) | −0.09*** | −0.07*** | −0.06*** |
| | (0.02) | (0.02) | (0.02) |
| Tenure | 0.41 | 0.54 | 0.24 |
| | (0.79) | (0.76) | (0.63) |
| $\text{Tenure}^2$ | −0.22** | −0.24** | −0.17** |
| | (0.08) | (0.07) | (0.06) |
| Skill (Elo, z-scored) | 0.15*** | 0.19*** | 0.25*** |
| | (0.03) | (0.03) | (0.03) |
| Opponent Skill (Elo, z-scored) | −0.16** | −0.20*** | −0.25*** |
| | (0.05) | (0.06) | (0.06) |
| Num. obs. | 334642 | 401913 | 496563 |
| $R^2$ (full model) | 0.32 | 0.30 | 0.27 |
| Adj. $R^2$ (full model) | 0.20 | 0.19 | 0.18 |
| Num. groups: human | 52179 | 52179 | 52179 |
| Num. groups: Year | 6 | 6 | 6 |
| Num. groups: Month | 12 | 12 | 12 |
| Num. groups: BotFESimple | 59 | 59 | 59 |
| Num. groups: GameSpeed | 6 | 6 | 6 |

$^{***}p < 0.001$; $^{**}p < 0.01$; $^{*}p < 0.05$; $^{\dagger}p < 0.1$

## First-Stage Regression of Endogeneity Correction

Our variables of interest, seeking AI feedback on games that were lost, gives rise to potential endogeneity concerns due to omitted variables as losing games itself is endogenous (i.e., individuals who experience more (less) failure, have more (less) opportunities to seek AI feedback
on those failures). The panel data structure and fixed effects specification underlying our analysis already accounts for measured and unmeasured individual-specific characteristics that are fixed over time. This includes important individual-level traits such as latent levels of ambition and learning goal orientation. Despite this approach, experiencing failure may potentially be endogenous due to time-variant omitted variables. We conduct a robustness test using a control function approach (Wooldridge, 2010). Table A2 below shows the results of the first-stage regression.

Contrary to two-stage least squares (2SLS) models, the control function approach does not allow for the direct test of relevance and exogeneity. However, we note that the instruments used in the first-stage are strong predictors of the (potentially) endogenous variable. For example, playing against a (perceived) stronger opponent increases the likelihood of losing the focal game ($\beta = 0.44$; $p < 0.001$). Furthermore, after periods with more (fewer) lost games, individuals are less (more) likely to lose the focal game (this is the opposite sign than in the pooled data, where prior losses are positively correlated with losing the focal game clearly indicating individual-level trait of ambitiousness).

We estimate first-stage models to predict the likelihood that human players would lose a game and then include the residuals from that regression as additional control variables in our main model estimating the likelihood to seek AI feedback (second-stage). There, the residuals represent the component of having lost the game that likely correlates with the error term. This approach ensures consistent second-stage estimates by directly controlling for the endogenous component of the problematic variable (Wooldridge, 2015). We bootstrap standard errors for the second-stage based on 1,000 replications, randomly drawing individuals (rather than observations) to retain the panel data structure. We then compute percentile bootstrap *p*-values. This approach is non-parametric and does not rely on asymptotic normality, making it well-suited to complex multi-stage estimators. Even when standard errors are large, the bootstrap distribution may be tightly concentrated away from zero, resulting in small *p*-values. This reflects consistent directional effects rather than unusually precise estimates.

As shown in the main text, including the control function residuals in the second-stage (Table 2, Models 3 and 4), changes our results substantially. The likelihood to seek AI feedback on losses changes from negative to positive. This indicates a negative bias, where time-variant shocks that promote losses (e.g., lower motivation) are correlated with lower likelihood to seek AI feedback. Accounting for the endogeneity in motivation, individuals do not seem to have an aversion to seeking AI feedback after experiencing failure. Taking the substantive results, and the changes in coefficients due to control function approach together, our results indicate that using AI feedback correctly follows both exogenous (skill) and endogenous processes where higher skilled (and likely more motivated, more learning oriented individuals) are more likely to seek AI feedback in situations that were challenging for them.

Table A2. First-stage results for control function endogeneity correction.

| | First-Stage (binomial) | |
|---|---|---|
| Dependent Variable: | Loss | AI Feedback |
| | (1) | (2) |
| Losses in Previous 5 Games | $-0.05^{***}$ | |
| | (0.01) | |
| Time Between Previous 5 Games (Percentile) | $-0.13^{**}$ | |
| | (0.04) | |
| New User (< 5 previous games) | $-0.20^{***}$ | |
| | (0.03) | |
| Opponent Skill | $0.44^{***}$ | |
| | (0.03) | |
| Aware of AI Feature | | $36.53^{***}$ |
| | | (0.54) |
| Same Day Peer AI Usage | | $0.04^{*}$ |
| | | (0.02) |
| Controls | | |
| Skill (Elo, z-scored) | $-0.08^{***}$ | $-0.07^{*}$ |
| | (0.02) | (0.03) |
| Experience (log) | $-0.03^{**}$ | $-0.77^{***}$ |
| | (0.01) | (0.02) |
| Tenure | $-0.03$ | 0.05 |
| | (0.30) | (0.63) |
| Tenure$^2$ | $0.12^{***}$ | 0.05 |
| | (0.04) | (0.09) |
| Common Opening | $-0.10^{***}$ | 0.02 |
| | (0.01) | (0.03) |
| Game Length | $-0.00$ | 0.00 |
| | (0.00) | (0.00) |
| Num. obs. | 336,477 | 187,422 |
| Num. groups: human | 21,743 | 10,164 |
| Num. groups: BotFESimple | 59 | 58 |
| Num. groups: Year | 6 | 6 |
| Num. groups: Month | 12 | 12 |
| Num. groups: GameSpeed | 6 | 6 |
| Num. groups: DayOfWeek | 7 | 7 |
| Deviance | 365,033.24 | 121,958.15 |
| Log Likelihood | $-182,516.62$ | $-60,979.07$ |
| Pseudo R$^2$ | 0.12 | 0.35 |

$^{***}p < 0.001$; $^{**}p < 0.01$; $^{*}p < 0.05$; $^{\dagger}p < 0.1$

## Robustness Test: Performance Improvements when Playing against Human Players

It is possible that using an AI to analyze a game only benefits a player who is playing against AI, and that the benefit of analyzing a game vanishes when playing against a human. To test this possibility, we randomly sampled 3,000 players who played at least one game against a bot on Lichess. We then gathered every game a player played against a human and calculated the accuracy for every move in every game the humans played against other humans, exactly like our analysis against bots. Unfortunately, we cannot observe when a human analyzes a game against another human on Lichess - Lichess will only report that a game was analyzed, not who analyzed it. However, we can still observe when a human analyzes a game against a bot, which are our focal variables below. We then limited our sample to the first fifty games against humans that a player played against, like in our prior analyses.

Table A3. Performance against human players.

| Dependent Variable: | Performance (OLS) | | | |
|---|---|---|---|---|
| | Model 1 | Model 2 | Model 3 | Model 4 |
| | (1) | (2) | (3) | (4) |
| Total Games Analyzed | 0.02 | | | |
| | (0.02) | | | |
| Losses Analyzed (lagged) | | 0.09 | | 0.15* |
| | | (0.06) | | (0.07) |
| Wins Analyzed (lagged) | | | 0.00 | −0.07 |
| | | | (0.04) | (0.05) |
| Controls | | | | |
| Human Games (log) | −0.01 | −0.03 | 0.02 | −0.01 |
| | (0.11) | (0.10) | (0.10) | (0.11) |
| Bot Games (log) | −0.00 | −0.00 | −0.00 | −0.00 |
| | (0.03) | (0.03) | (0.03) | (0.03) |
| Tenure | 1.51 | 1.51 | 1.51 | 1.48 |
| | (1.38) | (1.38) | (1.38) | (1.39) |
| Tenure$^2$ | −0.11* | −0.11* | −0.11* | −0.11* |
| | (0.05) | (0.05) | (0.05) | (0.05) |
| Skill (Elo, z-scored) | 0.15* | 0.15* | 0.15* | 0.15* |
| | (0.07) | (0.07) | (0.07) | (0.07) |
| Opponent Skill (Elo, z-scored) | −0.16** | −0.16** | −0.16** | −0.16** |
| | (0.06) | (0.06) | (0.06) | (0.06) |
| Year-Quarter FE | Yes | Yes | Yes | Yes |
| Individual FE | Yes | Yes | Yes | Yes |
| Num. obs | 105,423 | 105,423 | 105,423 | 105,423 |
| Num. Time Controls | 6 | 6 | 6 | 6 |
| $R^2$ (full model) | 0.22 | 0.22 | 0.22 | 0.22 |
| Adj. $R^2$ (full model) | 0.20 | 0.20 | 0.20 | 0.20 |
| Num. groups: UserName | 2459 | 2459 | 2459 | 2459 |

$^{***}p < 0.001$; $^{**}p < 0.01$; $^{*}p < 0.05$

Using panel data, we find that when humans analyze their wins against bots, they do not improve in future games (Model 4, $\beta = -0.07$, $p > 0.05$). However, just like our previous analyses showed, when humans analyze their losses against bots, we find a positive and significant effect that humans improve in accuracy in future games (Model 4, $\beta = 0.15$, $p < 0.05$). Our analysis indicates that analyzing a single loss improves future accuracy the same amount as increasing a player's Elo by a standard deviation.

# AI Feedback Leads to Specialization

This section provides additional details for our analysis of individual-level specialization. As mentioned in the main text, we focus on the opening moves of a chess game, known as "chess openings." Openings are important in chess because they set the stage for the middle game, thus narrowing the space of strategies that need to be considered in later phases of the game. They are widely studied and cataloged. Many opening sequences have well-known names such as the "Queen's Gambit" and the "Sicilian Defense" (Hooper and Whyld, 1992). Most players specialize in certain openings, focusing on a repertoire of set openings that they favor (Chassy and Gobet, 2011). While a narrow repertoire allows for deeper specialization it also makes players less flexible to deal with different opponents (Webb, 2013). The dependent variable is set to 1 if a player reuses an opening strategy that they have played recently (at least once in the previous five games), and 0 otherwise. The main explanatory variable in this panel regression analysis is the volume of AI feedback players sought between games. We estimate the following regression equation to investigate individual-level specialization:

$$Pr(RepeatStrategy_{it} = 1) = logit^{-1}(\beta_1 log(Cumulative\ AI\ Feedback_{it-1} + 0.5) +$$

$$\beta_2 Tenure_{it} + \beta_3 Tenure^2_{it} +$$

$$controls + \alpha_i + \alpha_{bot} + \alpha_{year} + \alpha_{month} + \alpha_{GameType} + \epsilon_{it})$$

Since moving first gives players more control over the direction of the game, we include an additional dummy indicating whether the focal player moved first (playing white) or second (playing black, omitted category) as an important control variable.

Table A4. Learning through specialization. Likelihood to repeat the same opening strategy increases with AI feedback.

| Dependent Variable: | Repeat Strategy (binomial) | |
|---|---|---|
| | Main Effect | Interaction |
| | (1) | (2) |
| Total AI Feedback (log) | 0.04* | 0.04* |
| | (0.02) | (0.02) |
| Skill (Elo, z-scored) | 0.01 | −0.00 |
| | (0.02) | (0.02) |
| Skill × Total Games Analyzed | | 0.02* |
| | | (0.01) |
| Controls | | |
| Playing White | 0.27*** | 0.27*** |
| | (0.02) | (0.02) |
| Common Opening | 1.19*** | 1.19*** |
| | (0.04) | (0.04) |
| Game Length | 0.00 | 0.00 |
| | (0.00) | (0.00) |
| Experience (log) | 0.02 | 0.03 |
| | (0.02) | (0.02) |
| Tenure | −1.42*** | −1.41*** |
| | (0.34) | (0.34) |
| Tenure$^2$ | 0.16*** | 0.16*** |
| | (0.04) | (0.04) |
| Opponent Skill | 0.03 | 0.03 |
| | (0.03) | (0.03) |
| Num. obs. | 238,987 | 238,987 |
| Num. groups: Individual | 11,239 | 11,239 |
| Num. groups: BotF | 59 | 59 |
| Num. groups: Year | 6 | 6 |
| Num. groups: Month | 12 | 12 |
| Num. groups: Game Type | 6 | 6 |
| Deviance | 246,930.11 | 246,922.54 |
| Log Likelihood | −123,465.05 | −123,461.27 |
| Pseudo $R^2$ | 0.07 | 0.07 |

$^{***}p < 0.001$; $^{**}p < 0.01$; $^{*}p < 0.05$; $^{\dagger}p < 0.1$